\documentclass[a4paper,10pt]{article}
\usepackage[utf8x]{inputenc}
\usepackage{threeparttable}
\pdfoutput=1
\usepackage{verbatim}
\usepackage{epsfig}
\usepackage{epsf}
\usepackage{epstopdf}
\usepackage{graphicx}
\usepackage{textcomp}
 \usepackage{amsmath}
\usepackage{amssymb}
\usepackage{color}
\definecolor{Blue}{rgb}{0.3,0.3,0.9}
\definecolor{red}{rgb}{1,0,0}
\usepackage{rotating}
\usepackage{pgfplots}
\usepackage{tikz}
\usepackage{url}
\usetikzlibrary{patterns}
\usepackage{amsfonts,amsthm,titling,url,array,mathtools}
\usepackage[superscript,biblabel]{cite}
\usepackage[toc,acronym]{glossaries}

\title{Mobile Artificial Intelligence Technology for Detecting Macula Edema and Subretinal Fluid on OCT Scans: Initial Results from the DATUM alpha Study}
\author{\hspace{-0pt}Stephen G. Odaibo, M.D., M.S.(Math), M.S.(Comp. Sci.)\textsuperscript{1},\\\hspace{-15pt} Mikelson MomPremier, M.D., FACS\textsuperscript{2},
Richard Y. Hwang, M.D., PhD\textsuperscript{3},\\ Salman J. Yousuf, D.O.,M.S.\textsuperscript{4}, Steven L. Williams, M.D.\textsuperscript{5},\\ Joshua Grant, M.D.\textsuperscript{6}}

\begin{document}

\maketitle

\begin{abstract}
\textbf{Importance:} Artificial Intelligence (AI) is necessary to address the large and growing deficit in retina and healthcare access globally. Mobile AI diagnostic platforms running
in the Cloud may effectively and efficiently distribute such AI capability.\\

\textbf{Objective:} To evaluate the feasibility of Cloud-based mobile artificial intelligence for detection of retinal disease. And to evaluate the accuracy of a particular such system
for detection of subretinal fluid (SRF) and macula edema (ME) on OCT scans.\\

\textbf{Design:} A multicenter retrospective image analysis was conducted in which board-certified ophthalmologists with fellowship training in retina evaluated OCT images of the macula.
They noted the presence or absence of macula edema or subretinal fluid, then compared their assessment to that obtained from Fluid Intelligence, a mobile AI app that detects SRF and ME on OCT scans.\\

\textbf{Setting:} Clinic setting with PACS access to OCT images of patients with or without retinal fluid.\\

\textbf{Participants:} Board-certified ophthalmologists with fellowship training in retina retrospectively selected retinal OCT images in a consecutive fashion. Effort was made by each center to balance
the number of scans with retinal fluid and scans without. Exclusion criteria included poor quality scans, ambiguous or overly subtle features, macula holes, retinoschisis, and dense epiretinal membranes. \\

\textbf{Main outcome and Measures:} Accuracy in the form of sensitivity and specificity of the AI mobile app compared against assessments of board-certified retina specialists.\\

\textbf{Results:} At the time of this submission, five centers have completed their initial studies. This consists of a total of 283 OCT scans of which 155 had either ME or SRF (“wet”) and 128 did not (“dry”).
The sensitivity ranged from 82.5\% to 97\% with a weighted average of 89.3\%. The specificity ranged from 52\% to 100\% with a weighted average of 81.23\%.\\

\textbf{Conclusion and Relevance:} Cloud-based Mobile artificial intelligence technology is feasible for the detection retinal disease. In particular, Fluid Intelligence (alpha version),
is sufficiently accurate as a screening tool for SRF and ME, especially in underserved areas. Further studies and technology development is indicated.\\

\vspace{250pt}

 \textbf{Author Affiliations:}\\\vspace{-10pt}
 \begin{flushleft}
\begin{itemize}
 \item[1.]\footnotesize RETINA-AI Health Inc., Houston TX,
 \item[2.]\footnotesize Southwest Retina Consultants, El Paso TX,
 \item[3.]\footnotesize MomPremier Eye Institute, Dallas TX,
 \item[4.]\footnotesize Saratoga Ophthalmology, Saratoga NY,
 \item[5.]\footnotesize Mid-South Retina Associates, Memphis TN,
 \item[6.]\footnotesize Bloomfield Eye Care, Bloomfield Hills, MI
\end{itemize}
\end{flushleft}

\vspace{20pt}
\textbf{Corresponding Author:}\\\vspace{-10pt}
\begin{itemize}
{ \item []\footnotesize \hspace{-10pt}Name: Stephen G.Odaibo, M.D.,M.S.(Math),M.S.(Comp. Sci.)}
 {\item[]\footnotesize \hspace{-10pt}Emial: stephen.odaibo@retina-ai.com}
\end{itemize}

\end{abstract}
\newpage
\section{Introduction}

Artificial intelligence will play a central role in screening of retinal diseases. And mobile devices are an important modality through which diagnostic capability can be distributed via the Cloud to areas of need.
Currently it is estimated that there are under 3000 retina specialists in the U.S. for its population of approximately 329.25 million people as of July 1st 2018 (U.S. Census Bureau)\cite{ce2019}. Furthermore, the retina
specialists are clustered in urban centers, and are few and far between in the rural and non-coastal states. For instance, it is estimated that the entire state of Wyoming has only two retina specialists in residence.
And as a result of this uneven distribution of specialized care, there are several severely underserved areas of the U.S. The numbers are even more dire for most other countries of the world. For instance, it is
estimated that, Nigeria, Africa’s most populous country has less than 10 fellowship-trained retina specialists in practice for a population of 190 million people. While Bolivia has only one. This severe shortage
of retina care is a global problem that cannot be adequately solved by training more physicians alone. The Association of American Medical Colleges (AAMC) projects that the problem is worsening. A 2018 AAMC report\cite{ma2018}
projects a shortfall of up to 121,300 physicians in the U.S. by 2030, and up to 72,700 of that shortage could be specialists. Therefore AI is clearly a needed supplement to the retina care system and to the
healthcare system in general. 
One important problem to solve, however, is the means and mechanism via which AI-driven diagnostic capability will be distributed to the areas where it is most needed. Fluid Intelligence to our knowledge is the
first mobile AI app for for eye care providers. It is an assist device for diagnosing the presence or absence of macula edema or subretinal fluid. The mobile app platform provides a ubiquitously available modality
whose reach is global. And this app presents an opportunity to assess the feasibility of the Mobile/Cloud-based AI distribution method. Practically, the app user points a mobile device camera at a picture of a retinal
OCT displayed on a computer monitor or piece of paper, and then snaps a picture of that picture. The taken picture then goes into the cloud, the engine computes a diagnosis, and returns a response to the user.
The problem of determining the presence or absence of subretinal fluid or macula edema on OCT is an important one. Firstly, the presence of macula edema or subretinal fluid in the macula is essentially always indicative
of pathology. It is a canonical hallmark that underlies several common retinal conditions such as diabetic macula edema\cite{hepu1998,madu2003,hepu1995,otki1999,puhe1995}, exudative macula degeneration
\cite{heba1996,husw1991,swiz1993,coco2007,kepa2012}, retinal vein occlusions \cite{rofu2005}, pseudophakic macula edema \cite{vooh2007,sosa1999,kibe2008,kieq2007,peut2007}
central serous chorioretinopathy \cite{hepu1995b,iiha2000,beka2000,moru2005,fugo2008}, and macula-off retinal detachments \cite{wogo2002,bahi2004,besc2007,wo2004,naha2009}. In non-obvious cases, the determination of retinal fluid is often a diagnostic dilemma for optometrists and even general ophthalmologists.
It constitutes a frequent source of unnecessary referral, or of non-referral where referral is indicated. In cases of unnecessary referral, the result is often unnecessary expensive and highly inconvenient long-distance travel
for elderly patients in underserved areas. It also results in over-crowding of retina physician offices by patients who need not be there, and this limits the capacity of retina physicians to see patients who do need to be seen.
In cases where referral is indicated but not done, the result is often outright vision loss.
In the remainder of this paper, we describe the DATUM study, a multicenter retrospective image analysis that tests the Fluid Intelligence mobile AI app for detecting the presence of fluid on OCT scans of the retina. Further,
we discuss study methodology, results, and implications.

\section{Methods}

The iOS operating system version of the Fluid Intelligence app was used on iPhone 6 devices. From within the app, the iPhone camera modality was used by study investigators to take pictures of OCT scans of the retina
displayed on computer monitors. The study investigators were board-certified ophthalmologists who have completed retina fellowships (“retina specialists”). The pictures of the OCTs were taken in anonymized fashion.
Study investigators selected images retrospectively and in consecutive fashion. Effort was made by each investigator to balance the number of OCT images with subretinal fluid (SRF) or macula edema (ME) (“wet”) against
images without SRF or ME (“dry”). Upon selecting an image, an assessment of the image was made by the investigator, after which the investigator used the app to determine the AI’s assessment. If the AI app’s assessment
of a “wet” image corresponded to the retina specialist’s, it was deemed a true positive. If the AI app was in agreement with the retina specialist on a “dry” image, it was deemed a true negative. If the AI app assessed
an image as “wet” but the retina specialist assessed it as “dry,” it was deemed a false positive. And if the AI app assessed an image as “dry” but the retina specialist assessed it as “wet,” it was deemed a false negative.
Investigators were at liberty to select the images provided they were in accordance with exclusion criteria. The following types of OCT images were excluded from the study: 1) poor quality images, 2) macula holes, 3)
dense epiretinal membranes, 4) macula retinoschisis, 5) vitreomacular traction syndrome, 6) and subtle or ambiguous images. Anonymized images were stored in the cloud and assessed for compliance with eligibility criteria.
An image eligibility check was done on all the centers’ submissions. 

\begin{figure}[h]
\begin{center}
\scalebox{.50}
{\includegraphics{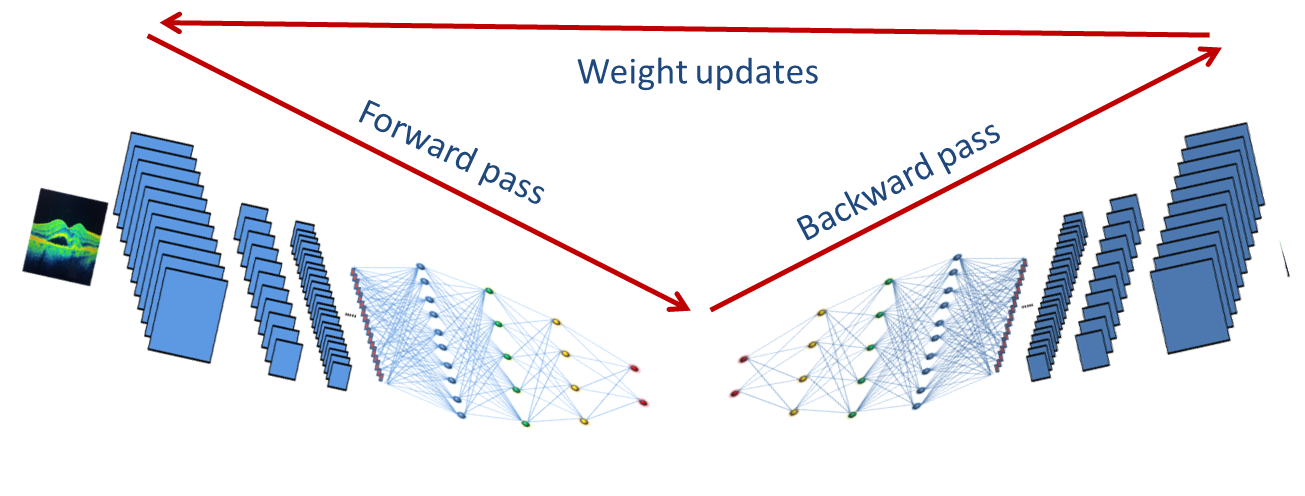}}
\end{center}
\caption[ML Procedure]{Machine Learning Procedure}
\label{fig:ML algorithm}
\end{figure}

Building the mobile AI app device: A model was trained on a data set consisting of 1 million OCT images after augmentation. The training was done with a deep learning model with an imageNet-trained inception base,
such as illustrated in Figure 1. Transfer learning was completed after which the trained model was hosted in the cloud. A front-end app for iOS was developed and deployed into the App Store. The front-end was connected
to the cloud service to serve the trained machine learning model during inference. An intermediate noSQL data base archived the images sent to the cloud server for inference.

\section{Results}

At the time of this submission, five centers had completed their initial studies. This consists of a total of 283 OCT scans of which 155 had either ME or SRF (“wet”) and 128 did not (“dry”). The sensitivity ranged
from 82.5\% to 97\% with a weighted average of 89.3\%. The specificity ranged from 52\% to 100\% with a weighted average of 81.23\%. Table 1 and Chart 1 show sensitivities and specificities of individual centers.

\begin{center}
\begin{table}[h] \caption{DATUM $\alpha$ Study Results}
\centering
\small{
\begin{tabular}{|c||c|c|c|c|}
\hline 
 & \textbf{Sensitivity} & \textbf{Specificity} & \textbf{Number wet} & \textbf{Number dry}  \\ [0.5ex]
\hline\hline 
\textbf{Center A} & 82.5\% & 66\% & 31 & 15\\\hline
\textbf{Center B} & 94\% & 52\% & 25 & 25\\\hline
\textbf{Center C} & 83.3\% & 100\% & 24 & 25\\\hline
\textbf{Center D} & 97\% & 100\% & 35 & 35\\\hline
\textbf{Center E} & 82.5\% & 75\% & 40 & 28\\
\hline
\end{tabular}
}
 \label{Tab:DATUM_alpha}
\end{table}
\end{center}

\begin{center}
    \begin{tikzpicture}
        \begin{axis}[ybar = 0.6,
width=1.0\textwidth,
height=0.4\textheight,
bar width=16pt,
ybar = 0.6,
ymin=0,
ymax=110,
symbolic x coords={Center A, Center B, Center C, Center D, Center E},
xtick=data,
title={DATUM $\alpha$ Study Results},
ylabel={Percentage (\%)},
          ]
            \addplot[ybar,fill=none, pattern=north east lines] coordinates {
                (Center A,   82.5)
                (Center B,  94)
                (Center C,   83.3)
                (Center D, 97)
                (Center E, 82.5)
            };
            
             \addplot[ybar,fill=none, pattern=dots] coordinates {
                (Center A,   66)
                (Center B,  52)
                (Center C,   100)
                (Center D, 100)
                (Center E,75)
            };
            \legend{Sensitivity, Specificity}
        \end{axis}
    \end{tikzpicture}
    \end{center}
    
    \section{Discussion}
    
The severe and growing shortage of retina care access in the U.S. and globally makes a compelling case for artificial intelligence driven supplementation and enhancement of care. However, discovering and developing
the most effective and efficient mechanisms by which AI-driven diagnostic capability will be distributed is essential. A mobile AI app modality could be both an effective and efficient means of distributing AI diagnostic capability.
Such a system, the first of its kind, Fluid Intelligence, was validated here in the DATUM alpha study. The results demonstrate that mobile AI running in the Cloud is an effective and efficient means of distributing AI diagnostic capability.
The DATUM alpha study initial results demonstrate a high enough sensitivity, 89.3\%, to make for a screening tool. Prior to the study, one could be justifiably skeptical that taking a picture of an OCT displayed on a computer
monitor would yield an image of sufficient quality to make an accurate AI diagnosis. The current study is named the alpha study as it is first in the series. As the machine learning algorithm is further trained with more data,
subsequent validation studies will be accordingly named beta, gamma, delta, etc. One of the advantages of running the machine learning algorithm in the cloud is the ability for periodically enhancing the algorithm’s training
and updating the AI seamlessly.
The sensitivities found on the DATUM alpha study demonstrated a tight band of 14.5 percentage points (from 82.5\% to 97\%) between centers. In addition to the reasonably high sensitivities across the board, the relatively
narrow span adds to the confidence that the AI app can satisfactorily detect a problem when there is one. In contrast to the sensitivities, the specificities had a rather wide range, spanning 48 percentage points (from 52\% to 100\%).
The exact specificities were 52\%, 66\%, 75\%, 100\%, and 100\%. The wide span of the specificities is likely explained by image selection. The algorithm was more likely to correctly identify an image as “dry” if the image was either
normal or had only moderate structural pathology. Also if an image was grainy due to low resolution but was still deemed to be of sufficient image quality to be eligible, such an image was more likely to yield a false positive on the AI read.
Furthermore, we observed that False positives were more likely in images that had features commonly associated with exudative macular degeneration; features such as large confluent druse or pigment epithelial detachments. These false positives
can therefore be ‘trained out’ with increase in the size and feature diversity of our training data set.

\begin{figure}[h]
\begin{center}
\scalebox{.50}
{\includegraphics{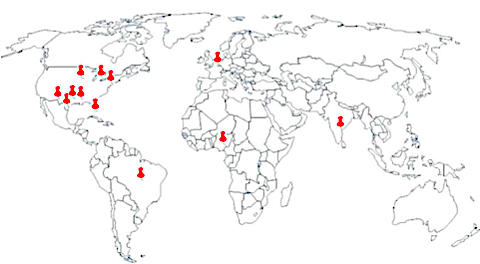}}
\end{center}
\caption[DSG]{DATUM Study Group}
\label{fig:DSG}
\end{figure}
    
Of note, the DATUM study group (DSG) is international with centers in over 15 countries and growing. Though all investigators in this initial submission are based in the U.S., subsequent studies will have more international representation
reflective of the DSG’s make up. Figure 2 shows the geographic location of DATUM study group investigators.

\section{Conclusion}

Cloud-based Mobile artificial intelligence technology is feasible for the detection of retinal disease. In particular, Fluid Intelligence (alpha version), is sufficiently accurate as a screening tool for subretinal fluid
and macula edema. This can serve to augment and retina care in underserved areas. Further studies and technology development are indicated; in particular, training on larger and more feature-diverse data sets.

\section*{Acknowledgement}
The authors are thankful to Dr. Michael H. Scott for kindly reviewing the draft and providing helpful suggestions. 

\bibliographystyle{plain}
\bibliography{/home/sodaibo/Documents/BOOKS/mybibliography_2015_nov}

\begin{thebibliography}{10}

\bibitem{bahi2004}
T.~Baba, A.~Hirose, M.~Moriyama, and M.~Mochizuki.
\newblock Tomographic image and visual recovery of acute macula-off
  rhegmatogenous retinal detachment.
\newblock {\em Graefe's Archive for Clinical and Experimental Ophthalmology},
  242(7):576--581, 2004.

\bibitem{beka2000}
T.~Bek and M.~Kandi.
\newblock Quantitative anomaloscopy and optical coherence tomography scanning
  in central serous chorioretinopathy.
\newblock {\em Acta Ophthalmologica Scandinavica}, 78(6):632--637, 2000.

\bibitem{besc2007}
S.~E. Benson, P.~G. Schlottmann, C.~Bunce, W.~Xing, and D.~G. Charteris.
\newblock Optical coherence tomography analysis of the macula after scleral
  buckle surgery for retinal detachment.
\newblock {\em Ophthalmology}, 114(1):108--112, 2007.

\bibitem{ce2019}
U.S.~Census Bureau.
\newblock U.s. census clock.
\newblock \url{https://www.census.gov/popclock/}, 2018 (accessed July 1st
  2018).

\bibitem{coco2007}
F.~Coscas, G.~Coscas, E.~Souied, S.~Tick, and G.~Soubrane.
\newblock Optical coherence tomography identification of occult choroidal
  neovascularization in age-related macular degeneration.
\newblock {\em American Journal of Ophthalmology}, 144(4):592--599, 2007.

\bibitem{fugo2008}
H.~Fujimoto, F.~Gomi, T.~Wakabayashi, M.~Sawa, M.~Tsujikawa, and Y.~Tano.
\newblock Morphologic changes in acute central serous chorioretinopathy
  evaluated by fourier-domain optical coherence tomography.
\newblock {\em Ophthalmology}, 115(9):1494--1500, 2008.

\bibitem{hepu1995b}
M.~R. Hee, C.~A. Puliafito, C.~Wong, E.~Reichel, J.~S. Duker, J.~S. Schuman,
  E.~A. Swanson, and J.~G. Fujimoto.
\newblock Optical coherence tomography of central serous chorioretinopathy.
\newblock {\em American Journal of Ophthalmology}, 120(1):65--74, 1995.

\bibitem{heba1996}
Michael~R Hee, Caroline~R Baumal, Carmen~A Puliafito, Jay~S Duker, Elias
  Reichel, Jason~R Wilkins, Jeffery~G Coker, Joel~S Schuman, Eric~A Swanson,
  and James~G Fujimoto.
\newblock Optical coherence tomography of age-related macular degeneration and
  choroidal neovascularization.
\newblock {\em Ophthalmology}, 103(8):1260--1270, 1996.

\bibitem{hepu1998}
Puliafito C. A. Duker J. S. Reichel E. Coker J. G. Wilkins J. R. Schuman J. S.
  Swanson E.~A. Hee, M.~R. and J.~G. Fujimoto.
\newblock Topography of diabetic macular edema with optical coherence
  tomography.
\newblock {\em Ophthalmology}, 105(2):360--370, 1998.

\bibitem{hepu1995}
Puliafito C. A. Wong C. Duker J. S. Reichel E. Rutledge B. Schuman J. S.
  Swanson E.~A. Hee, M.~R. and J.~G. Fujimoto.
\newblock Quantitative assessment of macular edema with optical coherence
  tomography.
\newblock {\em Archives of Ophthalmology}, 113(8):1019--1029, 1995.

\bibitem{husw1991}
David Huang, Eric~A Swanson, Charles~P Lin, Joel~S Schuman, William~G Stinson,
  Warren Chang, Michael~R Hee, Thomas Flotte, Kenton Gregory, Carmen~A
  Puliafito, et~al.
\newblock Optical coherence tomography.
\newblock {\em Science}, 254(5035):1178--1181, 1991.

\bibitem{iiha2000}
T.~Iida, N.~Hagimura, T.~Sato, and S.~Kishi.
\newblock Evaluation of central serous chorioretinopathy with optical coherence
  tomography.
\newblock {\em American Journal of Ophthalmology}, 129(1):16--20, 2000.

\bibitem{kepa2012}
P.~A. Keane, P.~J. Patel, S.~Liakopoulos, F.~M. Heussen, S.~R. Sadda, and
  A.~Tufail.
\newblock Evaluation of age-related macular degeneration with optical coherence
  tomography.
\newblock {\em Survey of ophthalmology}, 57(5):389--414, 2012.

\bibitem{kibe2008}
S.~J. Kim, M.~Belair, N.~M. Bressler, J.~P. Dunn, J.~E. Thorne, S.~R. Kedhar,
  and D.~A. Jabs.
\newblock A method of reporting macular edema after cataract surgery using
  optical coherence tomography.
\newblock {\em Retina}, 28(6):870--876, 2008.

\bibitem{kieq2007}
S.~J. Kim, R.~Equi, and N.~M. Bressler.
\newblock Analysis of macular edema after cataract surgery in patients with
  diabetes using optical coherence tomography.
\newblock {\em Ophthalmology}, 114(5):881--889, 2007.

\bibitem{ma2018}
I.~H.~S. Markit.
\newblock The complexities of physician supply and demand: Projections from
  2016 to 2030., 2018.

\bibitem{madu2003}
P.~Massin, G.~Duguid, A.~Erginay, B.~Haouchine, and A.~Gaudric.
\newblock Optical coherence tomography for evaluating diabetic macular edema
  before and after vitrectomy.
\newblock {\em American journal of ophthalmology}, 135(2):169--177, 2003.

\bibitem{moru2005}
J.~A. Montero and J.~M. Ruiz-Moreno.
\newblock Optical coherence tomography characterisation of idiopathic central
  serous chorioretinopathy.
\newblock {\em British Journal of Ophthalmology}, 89(5):562--564, 2005.

\bibitem{naha2009}
H.~Nakanishi, M.~Hangai, N.~Unoki, A.~Sakamoto, A.~Tsujikawa, M.~Kita, and
  N.~Yoshimura.
\newblock Spectral-domain optical coherence tomography imaging of the detached
  macula in rhegmatogenous retinal detachment.
\newblock {\em Retina}, 29(2):232--242, 2009.

\bibitem{otki1999}
T.~Otani, S.~Kishi, and Y.~Maruyama.
\newblock Patterns of diabetic macular edema with optical coherence tomography.
\newblock {\em American journal of ophthalmology}, 127(6):688--693, 1999.

\bibitem{peut2007}
I.~Perente, C.~A. Utine, C.~Ozturker, M.~Cakir, V.~Kaya, H.~Eren, Z.~Kapran,
  and O.~F. Yilmaz.
\newblock Evaluation of macular changes after uncomplicated phacoemulsification
  surgery by optical coherence tomography.
\newblock {\em Current Eye Research}, 32(3):241--247, 2007.

\bibitem{puhe1995}
C.~A. Puliafito, M.~R. Hee, C.~P. Lin, E.~Reichel, J.~S. Schuman, J.~S. Duker,
  J.~A. Izatt, E.~A. Swanson, and J.~G. Fujimoto.
\newblock Imaging of macular diseases with optical coherence tomography.
\newblock {\em Ophthalmology}, 102(2):217--229, 1995.

\bibitem{rofu2005}
P.~J. Rosenfeld, A.~E. Fung, and C.~A. Puliafito.
\newblock Optical coherence tomography findings after an intravitreal injection
  of bevacizumab (avastin{\textregistered}) for macular edema from central
  retinal vein occlusion.
\newblock {\em Ophthalmic Surgery, Lasers and Imaging Retina}, 36(4):336--339,
  2005.

\bibitem{sosa1999}
P.~Sourdille and P.~Santiago.
\newblock Optical coherence tomography of macularthickness after cataract
  surgery.
\newblock {\em Journal of Cataract \& Refractive Surgery}, 25(2):256--261,
  1999.

\bibitem{swiz1993}
E.~A. Swanson, J.~A. Izatt, M.~R. Hee, D.~Huang, C.~P. Lin, J.~S. Schuman,
  C.~A. Puliafito, and J.~G. Fujimoto.
\newblock In vivo retinal imaging by optical coherence tomography.
\newblock {\em Optics letters}, 18(21):1864--1866, 1993.

\bibitem{vooh2007}
B.~von Jagow, C.~Ohrloff, and T.~Kohnen.
\newblock Macular thickness after uneventful cataract surgery determined by
  optical coherence tomography.
\newblock {\em Graefe's Archive for Clinical and Experimental Ophthalmology},
  245(12):1765--1771, 2007.

\bibitem{wo2004}
T.~J. Wolfensberger.
\newblock Foveal reattachment after macula-off retinal detachment occurs faster
  after vitrectomy than after buckle surgery.
\newblock {\em Ophthalmology}, 111(7):1340--1343, 2004.

\bibitem{wogo2002}
T.~J. Wolfensberger and M.~Gonvers.
\newblock Optical coherence tomography in the evaluation of incomplete visual
  acuity recovery after macula-off retinal detachments.
\newblock {\em Graefe's Archive for Clinical and Experimental Ophthalmology},
  240(2):85--89, 2002.

\end{thebibliography}

\end{document}